\documentclass[12pt]{article}
\usepackage{graphicx,a4,aas_macros}   
\begin{document}

\newcommand{\preprintno}[1]
{{\normalsize\begin{flushright}#1\end{flushright}}}

\title{Axionic Dark Radiation and the Milky Way's Magnetic Field}
\author{Malcolm Fairbairn\thanks{E-mail address: malcolm.fairbairn@kcl.ac.uk}\\
{\em Theoretical Particle Physics and Cosmology, Department of Physics,}\\
{\em King's College London, The Strand, London WC2R 2LS, U.K.}}
\date{16th October 2013}

\maketitle
\begin{abstract}
Recently it has been suggested that dark radiation in the form of axions produced during the decay of string theory moduli fields could be responsible for the soft x-ray excess in galaxy clusters.  These soft X-ray photons come about due to the conversion of these axions into photons in the magnetic fields of the clusters.  In this work we calculate the conversion of axionic dark radiation into X-ray photons in the magnetic field of our own Galaxy.  We consider $\Delta N_\nu\sim 0.5$ worth of dark radiation made up of axions with energy of order 0.1-1 keV.  We show that it is possible, if a little optimistic, to explain the large regions of X-ray emission located above and below the centre of the Galactic plane detected in the 3/4 keV ROSAT all sky map completely due to the conversion of dark radiation into photons with an inverse axion-photon coupling of $M\sim 3\times 10^{13}$ GeV and an axion mass of $m\le 10^{-12}$ eV.  Different parameter values could explain both these features and the 3/4 keV X-ray background.  More conservatively, these X-ray observations are a good way to constrain such models of axionic dark radiation.
\end{abstract}
\section{Introduction}
The incredible view of the Cosmic Microwave Radiation provided by the Planck satellite has tightened our understanding of cosmology \cite{planck}.  One question it has not yet completely answered is whether there is an extra contribution to radiation other than the CMB photons themselves and the three known neutrino species.  While it has been pointed out that there is not strong evidence for requiring such radiation in a Bayesian sense \cite{hiranya} it is also true that there is still room for a small component of radiation.  In particular, discrepancies between the values of $H_0$ obtained by the HST and by Planck may be alleviated if a small component of dark radiation is added to the $\Lambda$CDM model (see \cite{steen} for a recent review).  Perhaps Planck's B-mode observations will be able to strengthen our understanding of the subject.  

One possible source of dark radiation would be string theory axions.  Axion like fields are often found in string theory compactifications \cite{witten,axiverse,bobbyaxiverse}, generically with an inverse coupling (corresponding to the Peccei Quinn Scale) of around $M\sim 10^{16}$ GeV, however, lower values of $M$ can be obtained depending upon the details of the compactification \cite{witten,andreas}.

String theory models often possess heavy moduli fields which need to decay in order not to dominate the energy density of the Universe during the radiation era.  The decay of the moduli fields would be gravitational and therefore democratic into every available degree of freedom but the weak coupling of the axion fields to the rest of the standard model means they will not come into thermal equilibrium and their temperature will not be related to the quarter power of their energy density.  More precisely, the typical energies of these axions will be greater than that of the CMB by a factor of $(M_{Pl}/M_\Phi)^{1/2}$ where $M_\Phi\sim 10^6$ GeV is the typical mass of the moduli fields \cite{adam,bobbyscott,bobby12}.  The energy of such dark radiation axions today would therefore be around 100 eV to a keV and can form a significant contribution to the energy density of dark radiation \cite{cicolijoe,higaki,joedod1}.

If those extra relativistic degrees of freedom are axion like particles then there is a non-zero probability for them to oscillate into photons in the presence of a magnetic field \cite{raffelt}.  If this is a truly relativistic form of dark radiation then it will form a flux of particles coming through the Galaxy from all directions on the sky.  Dark radiation axions arriving in the Galaxy can therefore convert into photons in the presence of the galactic magnetic field \cite{hoopandserp,brun,FRTgal}.  The purpose of this work is to quantify this effect.  Related recent work considers the conversion of dark radiation into photons in the primordial magnetic field \cite{higaki2,tashiro}.

It has been suggested recently that the excess soft X-ray radiation of galaxy clusters which has a badly understood origin \cite{softexcess} could be explained by string theoretic dark radiation axions converting into photons in the cluster magnetic field \cite{joedod2}.  The same authors suggested that the conversion of this axionic dark radiation in intergalactic magnetic fields might also explain part of the diffuse cosmic X-ray background radiation observed in the 3/4 keV band of the ROSAT satellite \cite{rosatallsky}.

In the next section we will describe the expected flux of dark radiation axions and the cosmic X-ray background that we are arguing could be explained by them.  Then we will describe the model of the Galactic magnetic field and distribution of dust - in order to do the mixing calculation we will need to know the distribution of hydrogen to find out the X-ray opacity and the electron density to understand the refractive index for photons which will affect their mixing.  We will then go onto describe the calculation of axion-photon mixing in this model of the Galaxy before moving onto presenting and then discussing our results.

\section{The Axion dark Radiation field and the Cosmic X-ray Background}
\subsection{Dark Radiation Flux}
The total energy density in the form of relativistic degrees of freedom (i.e. stuff that redshifts like radiation) can be expressed in the following form
\begin{equation}
\rho_{normal-rad}=\frac{\pi^2}{15}T_\gamma^4\left[1+\frac{7}{8}N_\nu\left(\frac{4}{11}\right)^{4/3}\right]
\end{equation}
where $N_\nu$ is the number of effective ``neutrino'' degrees of freedom.  In the standard model there are three neutrinos which due to a slight distortion of their thermal spectrum after they freeze-out leads to a value of $N_\nu=3.046$.  Extra relativistic degrees of freedom of any kind which make up the possible purported dark radiation mentioned earlier are usually expressed using this notation, as if we imagine they are neutrinos.  The relationship between the energy density of this dark radiation and the extra degree of freedom is therefore 
\begin{equation}
\rho_{dark-rad}=\frac{7}{8}\left(\frac{4}{11}\right)^{4/3}\Delta N_\nu\rho_\gamma=5.9\times 10^{-2}\Delta N_{\nu}{\rm eV cm^{-3}}.
\end{equation}
In this work we will be considering dark radiation axions converting into photons in the Galactic magnetic field.  The specific flux in units of ${\rm cm^{-2}sr^{-1}eV^{-1}s^{-1}}$ of such photons arriving in the Solar system would be given by
\begin{equation}
\frac{d\Phi_\gamma(E)}{dE}=P_{a\rightarrow\gamma}(E,l,b)\frac{d\Phi_a(E)}{dE}
\end{equation}
where $P_{a\rightarrow\gamma}(E,l,b)$ is the probability of an axion of energy $E$ arriving from outside the Galaxy from the direction $l,b$ in Galactic coordinates of converting into a photon of the same energy.
\subsection{Mixing probability required to explain Diffuse X-ray background at 3/4 keV \label{diffuseprob}}
One of the things we will be attempting to see in this work is if dark radiation could be responsible for the cosmic X-ray background.  This background radiation is only cleanly visible close to the poles of the Galaxy since it seems to reside in the 0.65-1keV energy range where there is a lot of emission from the Galactic disk

The spectrum of axions due to Moduli decay has to be obtained numerically as it depends upon how radiation or matter dominated the Universe is during the decay \cite{joedod2}. We assume that the situation we are imagining is similar to that envisaged in reference \cite{joedod2} where the Universe is matter dominated until the decay of the moduli in which case too a good degree of approximation the spectrum would take the form
\begin{equation}
\frac{d\Phi}{dE}\simeq 64.5\sqrt{\frac{\Delta N_\nu}{0.5}\frac{E}{eV}}\exp\left[-\left(\frac{E}{357\rm eV}\right)^2\right]{\rm cm^{-2}s^{-1}sr^{-1}eV^{-1}}\label{spectrum}
\end{equation}
In the range between 650 eV and 1 keV where the 3/4 keV background lies, the flux will be $7.4\times 10^7$eV cm$^{-2}$sr$^{-1}$s$^{-1}$. which should be compared with the observed diffuse X-ray flux in the same interval which is $1.0\times 10^{-12}$ergs cm$^{-2}$s$^{-1}$degree$^{-2}$ or $2.0\times 10^3$eV cm$^{-2}$sr$^{-1}$s$^{-1}$.  We therefore need around $3\times 10^{-5}$ of the axions to convert into photons for this energy in order to understand the 3/4 keV background as being due to dark axions.

Note this spectrum, originally plotted in reference \cite{joedod1}, peaks around 200 eV to try and offer a very exciting possible solution to the soft X-ray galaxy cluster excess \cite{joedod2,softexcess}.  Because of this, it is natural to consider slightly different moduli masses with energy spectra which could peak in the 650-1keV region.  However, if we shift the spectrum slightly to optimise this simply by changing the value in the denominator and changing the prefactor to keep the overall normalisation we get
\begin{equation}
\frac{d\Phi}{dE}\simeq 5.58\sqrt{\frac{\Delta N_\nu}{0.5}\frac{E}{eV}}\exp\left[-\left(\frac{E}{950\rm eV}\right)^2\right]{\rm cm^{-2}s^{-1}sr^{-1}eV^{-1}}
\end{equation}
which actually doesn't change the energy in the 0.64-1keV band much, we find instead $9.2\times 10^7$eV$^{-1}$cm$^{-2}$sr$^{-1}$s$^{-1}$ in that band.  This tells us that the results are not extremely dependent upon the precise value of the reheat temperature into axions.
\subsection{Mixing Probability required to explain Diffuse X-ray Background at 1/4 keV}
In addition to the background observed in the 3/4 keV band of ROSAT, there is also evidence for diffuse background in the 1/4 keV (100 eV-284 eV) ROSAT band with a lower limit of $5.5\times 10^3$eV$^{-1}$cm$^{-2}$sr$^{-1}$s$^{-1}$ and an upper limit of  $12\times 10^3$eV$^{-1}$cm$^{-2}$sr$^{-1}$s$^{-1}$ \cite{cui}.  The integral of the spectrum (\ref{spectrum}) in that energy region is $9.3\times 10^7$eV$^{-1}$cm$^{-2}$sr$^{-1}$s$^{-1}$ so in those regions of the sky where the 1/4 background is measured (e.g. Ursa Major) we cannot have a conversion probability at that energy bigger than $1.3\times 10^{-4}$ while if we hope to explain this background with conversion of dark radiation axions into X-rays we would require a conversion probability greater than $5.9\times 10^{-5}$.

\subsection{Mixing Probability required to explain X-ray emission from Fermi Bubbles}

We will also be considering whether the dark radiation could be associated with the X-ray emission viewed above and below the plane of the Galaxy which is thought to be associated with the Fermi Bubbles \cite{rosatallsky,tracy}.  The region above the Galactic plane in this area is confused by a loop which is thought to be associated with a relatively nearby supernova remnant.  Below the Galactic plane things are clearer and at coordinates $(l,b)\sim(6,-20)$ there is a flux of $1.9\times 10^{-11}$ergs cm$^{-2}$s$^{-1}$degree$^{-2}$ in the 3/4 keV (R45) ROSAT band which goes from 0.47 keV to 1.21 keV \cite{rosatallsky,heasarc}, corresponding to 3.9$\times 10^{4}$eV s$^{-1}$cm$^{-2}$sr$^{-1}$.

If we were to imagine that this flux therefore comes from dark radiation axions converting into X-rays we would require a probability conversion in this region of order $4.4\times 10^{-4}$.

\section{Galactic Model}
In order to find out whether an axion will mix with a photon as it moves through the Galaxy, we require a model of the Galaxy in the first place.  In this section we will describe how we model the Galactic dust, magnetic field and electron density

\subsection{Large scale Galactic Magnetic Field}

An essential input for the calculation is of course the magnetic field of the Milky Way Galaxy.  For this we use the detailed model of \cite{magfield} which is based upon a combination of the WMAP7 Galactic Synchrotron emission map and many extragalactic Faraday Rotation measurements.  The rather detailed model is the most complete Galactic Magnetic field model available at the time of writing and contains a disk component which follows the Milky Way Spirals, a toroidal component and an out of plain 'X' component, as described in detail in reference \cite{magfield}. 

\subsection{Fermi Bubbles}
The Fermi bubbles are large roughly spherical structures a few kpc in size which are located above and below the Galactic plain in the centre of the Galaxy.  These lobes emit gamma rays and were originally detected in the Fermi data by the authors of \cite{tracy} although they have also been analysed by the Fermi team \cite{fermifermibub}.  It has been pointed out in \cite{tracy} that all-sky ROSAT observations in the 1990s also saw emission which could be associated with large scale features above and below the central region of the Galaxy.  It has become clear through radio observations that these objects also emit tremendous amounts of radio waves and modeling has shown that their internal magnetic fields are extremely large, or the order of 10 $\mu$G \cite{naturefermi}.  The origin of the bubbles is thought to be either due to the activity of the central black hole or due to star formation occurring at the Galactic centre.

There are three types of components associated with these bubbles which are visible in the radio part of the spectrum \cite{naturefermi}:- there are the lobes themselves which we model as spheres of radius 3.8 kpc on either side of the centre of the Galactic disk, touching the disk itself.  The radio synchrotron emission from these lobes can be interpreted as coming from a diffuse magnetic field filling the lobes of 6$\mu$G or from plasma trapped in a shell of thickness 0.3 kpc around the boundary of the lobes, in which case the magnetic field required to explain the emission is around 12$\mu$G.  We will consider both situations.

Then there are ridges of emission which are thought to be associated with historical periods of activity and a central spur which is the same size as the ridges but seems to be in contact with the centre of the Galaxy, making it natural to interpret this feature as being due to relatively recent events at the Galactic centre.  The physical size in terms of height and width of these features (i.e. both the ridges and the spurs) and their magnetic fields are roughly equivalent so for our model we will consider only one feature with a width of less than a kpc and a height of around 4 kpc which is a rotated lemniscate of Bernoulli centred at the core of the Galaxy.  We will assume the magnetic field inside this region is 14 $\mu$G following the values given in \cite{naturefermi}.

Our modeling of these features is highly idealised.  The actual radiation emission areas have been distorted by various winds to form axi-asymmetric features.  In our toy model everything is completely axisymmetric.  Nevertheless, it should give us some indication of the amount of conversion we expect to see in the actual features which are observed.
\subsection{Distribution of Hydrogen and free electrons\label{gas}}
As we will see in the next section, the opacity of the Milky Way to X-rays is normalised relative to the column density of neutral hydrogen.  We also need to know the number density of free electrons in order to calculate the effective mass for the photons moving through the medium, which will affect their mixing probability (this is equivalent to the MSW or matter effect in neutrino mixing).  The distribution of $H_2$, $H_I$ and free electrons in the Milky way is a complicated subject that can be studied at a very deep level but for the purposes of this work we will adopt simple axisymmetric models with north south symmetries.  We will simply note that there does seem to be evidence of North-South Asymmetry in the distribution of neutral hydrogen which could be incorporated into a more advanced analysis \cite{nsasymh2}.

We adopt a simple model based upon the effect of dust upon COBE/FIRAS observations \cite{misiriotis}.  The number density of $H_2$ and $H_I$ are given by
\begin{equation}
n_{H_2}(z,r)=4.06\quad{\rm cm^{-3}}\exp\left(-\frac{r}{2.57{\rm kpc}}-\frac{|z|}{\rm 0.08 kpc}\right)
\end{equation}
where $r$ here is the cylindrical radial coordinate
\begin{equation}
n_{H_I}(z,r) = \left\{ \begin{array}{ll}
         0.32\quad{\rm cm^{-3}}\exp\left(-\frac{r}{18.24{\rm kpc}}-\frac{|z|}{\rm 0.52 kpc}\right) & \mbox{if $\rho \geq 2.75$kpc};\\
        0 & \mbox{if $\rho < 2.75$ kpc}.\end{array} \right. 
\end{equation}
where $\rho=\sqrt{x^2+y^2}$ is the radial distance in Galactic cylindrical coordinates.

For the free electrons, we adapt the approach of \cite{BMM06,gaensler} and assume that the local electron density has a value of $2.13\times 10^{-2}$cm$^{-3}$ constant up to values of $|z|=1$kpc then drops off with a scale height of 500 pc above this.  We keep the same vertical scaling as a function of $\rho$ but change the central value so it scales with $n_{H_I}$
\begin{equation}
n_e(z,r)=2.13\times 10^{-2}{\rm cm^{-3}}\exp\left( -\frac{r-R_\odot}{18.24\rm kpc}\right){\rm min}\left[1,\exp\left(\frac{|z|-1{\rm kpc}}{0.5{\rm kpc}}\right)\right]
\end{equation}
and we note that the ambient electron density outside the Milky Way seems to be badly measured.  In clusters the intergalactic electron density is of order $10^{-3}$cm$^{-3}$ \cite{joedod2} but the local group is not a dense cluster and we assume that the electron density in the outer halo of the Milky Way is much lower.
\begin{figure}[!h]
\centering
\includegraphics[scale=0.5]{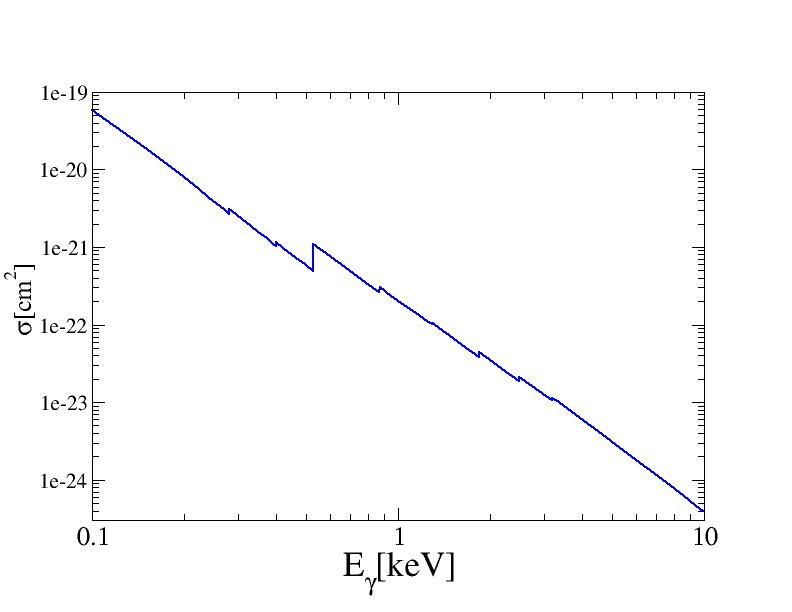}
\caption{\it Cross section per hydrogen atom for X-ray absorption in interstellar gas \cite{browngould}\label{opacfig}}
\end{figure}

\subsection{X-ray Opacity\label{opacsigma}}
We also need to take into account the fact that any X-rays which are produced will have to contend with the opacity of the Galaxy.  It is well known that the centre of the Galaxy is opaque to X-rays (see e.g. \cite{lydia}) and we will have to estimate this opacity by taking into account the cross-section for scattering of X-rays off atoms.  We use the paper of Brown and Gould to approximate the cross section for X-rays of different energy scattering off interstellar gas of solar abundance which is expressed as a cross section per hydrogen atom \cite{browngould}.  We then work out at every point along the path of the photon-axion trajectory the local density of hydrogen atoms using the model distributions of section \ref{gas} and obtain the mean free path for the photon parts of the density matrix (see section \ref{mixing}, in particular equation (\ref{damping}) ).  Figure \ref{opacfig} shows the opacity as a function of energy for X-ray absorption in interstellar gas.
\section{Mixing\label{mixing}}

The Lagrangian describing the photon and axion takes the following form,
\[
\mathcal{L}=\frac{1}{2}(\partial^\mu a\partial_\mu a-m^2 a^2)
-\frac{1}{4}\frac{a}{M}F_{\mu\nu}\widetilde
F^{\mu\nu}-\frac{1}{4}F_{\mu\nu}F^{\mu\nu},
\]
where $F_{\mu\nu}$ is the electromagnetic stress tensor and
$\widetilde
F_{\mu\nu}=\epsilon _{\mu \nu \rho \lambda }F_{\rho \lambda }$
is its dual,
$a$ denotes the pseudo-scalar axion, $m$ is the axion mass and $M$ is the
inverse axion-photon coupling. Because of the
$F_{\mu\nu}\tilde{F}^{\mu\nu}$ term, there is a finite probability for the
photon to mix with the axion in the presence of a magnetic field \cite{sikivie, raffelt}. We will be interested in light, $m
\le 10^{-13}$~eV, axions with inverse coupling mass scale $M$ between $10^{12}$~GeV and $10^{16}$ GeV.  The choice of the mass of the axions is because higher mass axions will lead to suppressed mixing, as we shall see presently.

We represent the photon
field $A(t,x)$ as a superposition of fixed-energy components
$A(x)e^{-i\omega t}$. If the magnetic field does not
change significantly on the photon wavelength scale and the index of refraction of the medium
$|n-1|\ll 1$, one can decompose~\cite{raffelt} the operators in the field
equations as (for a photon moving in the $z$ direction)
$\omega^2+\partial_z^2\rightarrow
2\omega(\omega-i\partial_z)$, so that the field equations become
Schrodinger-like,
\begin{equation}
i\partial_z \Psi=
-\left(\omega+\mathcal{M}\right)\Psi\qquad;\qquad
\Psi=\left(\begin{array}{c}A_{x}\\A_{y}\\a\end{array}\right),
\label{schrodinger}
\end{equation}
where
\[
{\cal M}\equiv\left(
\begin{array}{ccccccccc}
\Delta_p&0&\Delta_{Mx}\\
0&\Delta_p&\Delta_{My}\\
\Delta_{Mx}&\Delta_{My}&\Delta_m
\end{array}
\right).\hspace{0.7cm}
\]
for anything other than very high energy photons, he mixing is determined by the refraction parameter $\Delta_p$,
the axion-mass parameter $\Delta_m$ and the mixing parameter $\Delta_M$. The numerical values of these three parameters are 
\[
\begin{array}{rcccl}
\displaystyle \Delta_{M\! i} \!\!\!&=&\!\!\! \displaystyle \frac{B_i}{2M}
\!\!\!&=&\!\!\! \displaystyle
5.4\times10^{-5}
\left(\frac{B_i}{1\mu\mbox{G}}\right)
\left(\frac{10^{14}~\mbox{GeV}}{M}\right)\mbox{kpc}^{-1}\!\\
\displaystyle \Delta_m \!\!\!&=&\!\!\! \displaystyle \frac{m^2}{2
\omega} \!\!\!&=&\!\!\! \displaystyle 7.8\times 10^4\left(\frac{m}{10^{-9}~\mbox{eV}}\right)^2
\left(\frac{\mbox{keV}}{\omega}\right)\mbox{kpc}^{-1}\!\\
\displaystyle \Delta_p \!\!\!&=&\!\!\! \displaystyle \frac{\omega
_p^2}{2 \omega } \!\!\!&=&\!\!\! \displaystyle 1.1\!\!\left(\frac{n_e}{10^{-2}~\mbox{cm}^{-3}}\right)
\!\left(\frac{~\mbox{keV}}{\omega}\right)\mbox{kpc}^{-1}\!
\label{lengths}
\end{array}
\]
respectively.  Here $\omega _p^2=4\pi \alpha n_e/m_e$ is the plasma
frequency squared (effective photon mass squared), $n_e$ is the electron
density, $B_i$, $i=x,y$ are the the components of the magnetic field $B$ perpendicular to the direction of propagation,
~ $m_e$ is the electron mass, $\alpha $ is the fine-structure constant and
$\omega $ is the photon (axion) energy.

We integrate numerically the equations of motion
along a 30 kpc path starting far outside the Milky Way and ending at the location of the solar system, 8.5 kpc from the centre of the Galaxy. Instead of
explicitly solving the Schr\"{o}dinger equation, we go to the interaction representation and separate out the mixing that
we are interested in from the normal propagating oscillation.  This can be
done by defining a density matrix $\rho=\Psi^* \Psi$ with an evolution
equation \cite{edvard,FRTsun}
\begin{equation}
i\partial_z \rho=[\mathcal{M},\rho]-i{\cal D}\rho
\label{diffeq}
\end{equation}
where here we have simultaneously introduced a damping matrix $\cal{D}$
designed to take into account interactions between the photon part of the
wavefuction and the particles in the Galaxy.  This damping matrix takes
the form
\begin{equation}
{\cal D}\equiv\left(
\begin{array}{ccccccccc}
\Gamma&\Gamma&\frac{1}{2}\Gamma\\
\Gamma&\Gamma&\frac{1}{2}\Gamma\\
\frac{1}{2}\Gamma&\frac{1}{2}\Gamma&0
\end{array}
\right)
\label{damping}
\end{equation}
where the inverse decay length $\Gamma=n_{e}\sigma$.  We use cross section outlined in section \ref{opacsigma} and the gas density described in section \ref{gas}.
\begin{figure}[p]
\centering
\includegraphics[bb = 140 120 630 400,clip,scale=0.65]{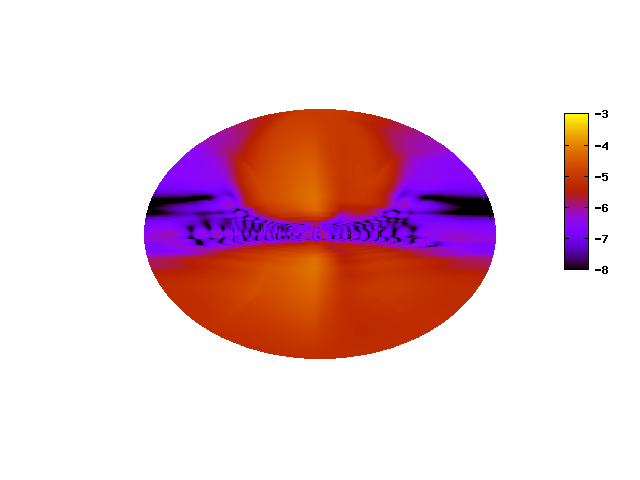} 
\caption{\it $\log_{10}$ of the probability of a photon arriving at Sun through normal Galactic field.  $M=10^{13}$GeV, $m=10^{-5}$neV, $\omega=800$eV.  Plot corresponds to Galactic coordinates with $(b,l)=(0,0)$ at the centre, $b$ increases vertically and $l$ increases to the right. \label{egg1}}
\includegraphics[bb = 140 120 630 400,clip,scale=0.65]{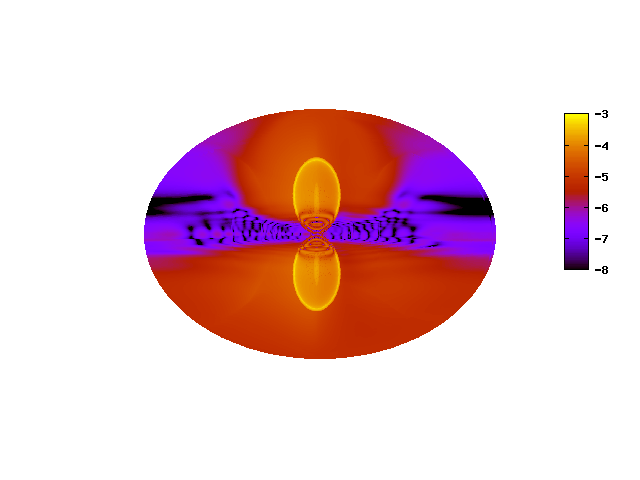}
\caption{\it Same as figure \ref{egg1} but with magnetic field corresponding to the Fermi bubble with shell configuration.\label{egg2}}
\includegraphics[bb = 140 120 630 400,clip,scale=0.65]{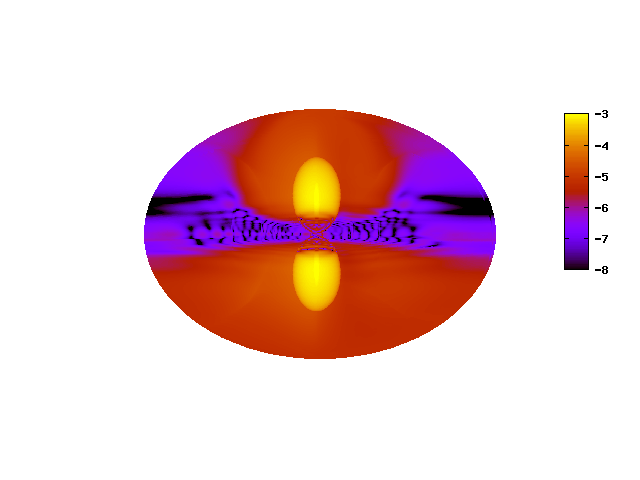}
\caption{\it Same as figure \ref{egg1} but with magnetic field corresponding to the Fermi bubble with full configuration.\label{egg3}}
\end{figure}
\section{Results}

Having established our model of the Galaxy, we consider the conversion of axions of energy 100 eV - 1 keV into photons in its magnetic field.  We evolve the equations set out in equation (\ref{diffeq}) through the Galaxy starting with a pure axion state and see what the probability of a photon arriving at Earth.

The basic illustration of the probability of a photon arriving is expressed most clearly in figures \ref{egg1},\ref{egg2} and \ref{egg3} which are all sky maps in Galactic coordinates of the probability of photons being produced as 800 eV axions move through the Galaxy for an inverse coupling of $M=10^{13}$GeV and $m=10^{-5}$neV (neV=$10^{-9}$eV).  It is clear that typical probabilities of $10^{-4}$ can be obtained without the Fermi bubbles rising to $10^{-3}$ when the Fermi bubbles are included.
\begin{figure}[!h]
\centering
\includegraphics[scale=0.5]{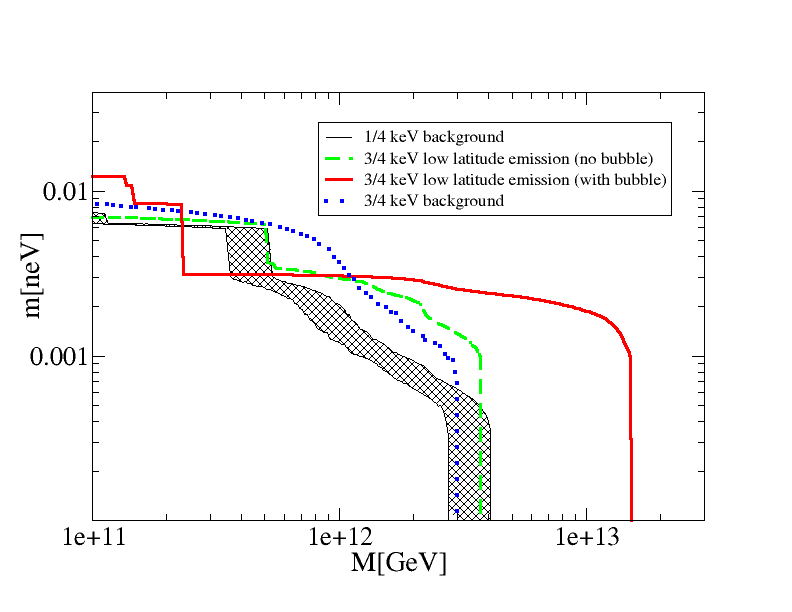} 
\caption{\it Values of the inverse coupling $M$ and the axion's mass $m$.  The black hatched band corresponds to the region of parameter space where one can explain the 1/4 keV emission as being due to axionic dark radiation converting into X-rays.  The Green dashed line corresponds to the parameters required to explain completely the low latitude emission above and below the centre of the Galaxy.  The red solid line is the same region assuming the Fermi bubbles are filled with a magnetic field of 6$\mu$G.  Finally the blue dotted line corresponds to explaining the 3/4 keV background radiation as being entirely due to dark radiation.  The step like features are discussed in text.\label{bigplot}}
\end{figure}
We are now in a position to test various hypothesis:-
\begin{itemize}
\item{Could these photons coming from axions explain the cosmic X-ray background? What value of the coupling and mass would lead to a signal in excess and therefore incompatible with the observed cosmic X-ray backgrounds at 1/4 and 3/4 keV?}
\item{Given the recently discovered high magnetic fields in the Fermi bubble, what combination of couplings and mass are acceptable such that emission from that region due to purported dark radiation is not in excess of the flux observed by ROSAT?}
\item{What are the implications of the answers to the previous two questions upon the possible explanation of the soft X-ray emission excess in galaxy clusters suggested in \cite{joedod2}}
\end{itemize}
In order to answer the first question, we first ask what set of axion parameters are required such that the flux of axions from dark radiation, upon converting into X-rays in the Galactic magnetic field, would have roughly the correct magnitude to explain the cosmic X-ray background.  We do this by noting in section \ref{diffuseprob} we found that (for $\Delta N_\nu=0.5$) we require a probability of mixing of $3\times 10^{-5}$ in order to explain the 3/4 keV cosmic X-ray background in this way.  We send axions through the north and south poles of the Galaxy where there would be less X-ray emission from objects in the Galaxy, although also less integrated magnetic field.  The parameters which lead to the appropriate probability are plotted in figure \ref{bigplot}.  As can be seen, there is a strong mass dependence, for too high axion mass, the coupling $1/M$ has to be increased rapidly, although for masses below $10^{-12}$eV the mass dependence disappears and we see that we require a value for $M$ which is around $10^{12}$GeV.

In this line, and in all lines, we can see various step-like features where as we increase the inverse coupling, the mass corresponding to the same probability drops almost vertically at particular points.  The origin of these features is because of the fact that the mixing lengths given by the equations (\ref{lengths}) are of the same order as the size of the domains in the Galactic model so by changing the lengths we can move between situations of constructive and destructive combined effects at the boundaries of different domains.

The then perform the same analysis to obtain the appropriate probability to explain the 1/4 keV X-ray background as being due to this dark radiation axion conversion.

Next we go on to look at what parameter values would be incompatible with the observed ROSAT emission observed above and below the poles of the Galaxy in roughly the same region as the Fermi Bubbles.  We choose a direction at (l,b)=(0,$-\pi/8$) which happens to be close to the maximum position for the probability of an X-ray photon emerging for our magnetic field configurations including bubbles and is also in the region where the ROSAT all sky survey observed high levels of X-ray emission outside the Galactic plane.  In the same figure \ref{bigplot} we show the values of $M$ and $m$ which could account for the ROSAT X-ray emission in this part of the sky.  There are two different results depending upon whether we assume that the observations suggesting very strong magnetic fields in the Fermi bubbles are correct or not.

\section{Discussion and Conclusions}
The results set out in figure \ref{bigplot} show that if dark radiation exists in the form of a sea of ultra light ($<10^{-12}$eV) axions with energies around a keV then there are constraints on the inverse coupling $M$ of these models due to conversion in our Galaxy which are typically a lot greater than in normal Axion Like Particle models (as one would expect given the very large number of axions required for them to form a significant dark radiation contribution).

At the point on figure \ref{bigplot} where $M\sim 10^{12}$ GeV and $m\sim 3\times 10^{-12}$eV one can completely explain both the 3/4 keV background radiation and the 3/4 keV radiation at low-latitude above and below the poles of the Galaxy as being entirely due to the conversion of dark radiation into X-rays rather than any astrophysical origin.  Clearly such an interpretation is rather extreme but it is nevertheless worth pointing out.  

Conversely, the 1/4 X-ray keV background cannot be explained because the parameter values which would allow one to explain this background as being due to dark radiation axions either leads to too large a 3/4 keV background when $m>3\times 10^{-12}$eV or too large low latitude 3/4 keV emission when $m<3\times 10^{-12}$eV (for the Fermi bubble magnetic field, the low latitude constraint emission would be the tightest constraint again above $m\sim 8\times 10^{-12}$eV).

We can however explain the low latitude emission in the ROSAT 3/4 keV maps above and below the Galactic centre with an inverse axion-photon coupling of $M\sim 1.5\times 10^{13}$ GeV and an axion mass of $m\le 10^{-12}$ eV without creating too much background in either the 1/4 keV or 3/4 keV bands thanks to the very large fields that have been observed in the Fermi bubbles.  In fact, it is obvious that the very large magnetic fields observed in the Fermi bubbles will have a very significant effect upon any conversion between axions and photons in the Galaxy in general.

The observation by the authors of \cite{joedod2} that the hard UV excess observed in Galaxy clusters can be explained by dark radiation requires values of $M\sim 10^{13}$GeV and $m<10^{-12}$eV.  There is a certain amount of freedom in these numbers depending upon the exact value of the magnetic fields in the clusters and the coherence of the field lengths.  We can see that these numbers would also produce interesting signals in the Milky Way if we were to better understand the nature of the diffuse X-ray emission, especially in the low latitude regions above and below the poles of the Galaxy.  Since the nature of the Fermi bubbles is so critical in understanding the photon-axion conversion in this part of the sky and since we are still learning about the nature of the Fermi bubbles, progress on understanding and strengthening these constraints could be made relatively quickly.

In summary, if the hard UV excess observed in galaxy clusters can be explained by dark radiation in the form of axions converting into X-ray photons in the magnetic fields of those clusters, we should also be able to see a signal from axion-photon conversion in the magnetic field of the Milky Way.  Since we can measure the magnetic field in our own Galaxy with some degree of accuracy, we can in this way obtain more reliable constraints upon these kind of dark radiation scenarios.

\section*{Acknowledgments}
It is a pleasure to thank Bobby Acharya, Joe Conlon and David Marsh for careful reading of a draft version of this paper and Sergei Troitsky and Timur Rashba for discussions in the past. Terry Sloan provided great help in checking the Galactic magnetic field calculation. MF is extremely grateful to recieve support from the STFC.

\end{document}